\documentclass[english]{article}
\usepackage[T1]{fontenc}
\usepackage[latin9]{inputenc}
\usepackage{geometry}
\usepackage{fancyhdr}
\usepackage{color}
\usepackage{textcomp}
\usepackage{amsmath}

\makeatletter
\usepackage{fancyhdr}
\usepackage{lastpage}
\makeatother

\usepackage{babel}
\begin{document}

\title{Non-spatial Probabilistic Condorcet Election Methodology}

\author{Ben  Wise%
\thanks{This work was performed in 2013 when the lead author was employed
at BAE Systems%
} {} and Steven Bankes, BAE Systems}

\date{7 December 2014}
\maketitle
\begin{abstract}
There is a class of models for pol/mil/econ bargaining and conflict
that is loosely based on the Median Voter Theorem which has been used
with great success for about 30 years. However, there are fundamental
mathematical limitations to these models. They apply to issues which
can be represented on a single one-dimensional continuum, like degree
of centralization of a government, or cartel members negotiating the
price to ask for their commodity. They represent fundamental group
decision process by a deterministic Condorcet Election. There has
been some extension to multidimensional issue sets, but they are not
well documented and are limited to cases where the difference between
policies is well approximated by a Euclidean distance and each actor's
utility is monotonically declining in distance. This work provides
a methodology for addressing a broader class of problems.

The first extension is to continuous issue sets where the consequences
of policies are not well-described by a distance measure or utility
is not monotonic in distance. Simple one-dimensional examples are
discussed. The difficulty is more acute in multidimensional policy
spaces. An example is the negotiations over national economic policies,
where the effects differ by region, by industrial sector, and by social
group. Further, even a weighted sum over the effects can be non-monotonic
or multi-peaked. 

The second fundamental extension is to inherently discrete issue sets.
The discussion will focus on subset selection problems, though the
methodology is not limited to them. Two examples are the selection
of which subset of competing parties will form a parliament (and which
will be excluded), or the selection of a portfolio of defense projects.
In the parliament formation case, the utility of a potential parliament
to each actor can be modeled as a look-ahead by that actor in order
to consider the various policies which that parliament might choose.
This models a two-stage process, where the uncertainty in the second
stage (choices on issues) is an important factor in the first stage
(choices of parliaments). Because there are generally more issues
than parties, the discrete choices in the first stage embody trade-offs
in the second stage. Because the options cannot easily be mapped into
a multidimensional space so that the utility depends on distance,
we refer to it as a non-spatial issue set.

The third, but most fundamental, extension is to represent the negotiating
process as a probabilistic Condorcet election. This provides the flexibility
to make the first two extensions possible; this flexibility comes
at the cost of less precise predictions and more complex validation.
Because the analyses are inherently probabilistic, this provides a
smooth \textquotedbl{}response surface\textquotedbl{} for expected
utility, thus simplifying strategy optimization even in discrete issue
sets. Some common bargaining algorithms are inapplicable in the more
general issue sets addressed here. For example, power-weighted interpolation
between the positions of two actors is often inapplicable in continuous
but multimodal issue sets. More fundamentally, interpolation is unusable
in discrete issue sets because continuous interpolation is not even
defined between discrete options.

We provide motivation and overview of the general methodology followed
by mathematical details. The methodology has been implemented in two
proof-of-concept prototypes which address the subset selection problem
of forming a parliament, and strategy optimization for one-dimensional
issues. Both prototypes can explore the effect of altering many parameters
and sub-models.

\end{abstract}

\section{Problem and State of the Art\label{sec:Problem-and-SOA}}

Keywords: Multi-agent Models; Collective Decision Making; Public Choice;
Bounded Rationality; Agent-Based Simulation; Voting Behavior; Social
Simulation

Many political, military, and/or economic disputes involve bargaining
and conflict between powerful groups. They can be modeled as thousands
to millions of small actors, or as a small number of larger actors;
we consider the latter class of models. Game theory, economics, and
public choice theory all play important roles in different contexts,
but one common theme is the use of models loosely based on the Median
Voter Theorem (MVT). There is a vast amount of academic analysis,
formal modeling tools , and popular political commentary based on
the same concepts, which we will collectively refer to as spatial
political models (SPM).%
\footnote{The ``spatial model of politics'' appears to have a somewhat more
restricted usage in the literature.%
} We present a methodology called Non-spatial Probabilistic Condorcet
Elections (NPCE) and describe some advantages and disadvantages compared
to SPM. This is not the only way to approach the problem of CE which
lack winners. Some approaches avoid the CE concept entirely, some
retain the idea of a CE but add some probabilistic elements. The usual
probabilistic approach is to model the CE as a deterministic function
with stochastic parameters. As explained below, our approach treats
the CE itself as stochastic, i.e. a stochastic function even with
deterministic parameters.

\cite{PolEconTPGT} gives a succinct motivation: \textquotedbl{}the
probabilistic voting approach was developed in the spatial voting
model to guarantee the existence of equilibrium in situations, such
as multidimensional policy space, in which a Condorcet winner fails
to exist; see Coughlin 1992 for an overview of probabilistic voting
and Osborne (1995) for an overview of spatial voting theory\textquotedbl{}
(the two citations are \cite{ProbVoteTheory} and \cite{SpatialModelsUnderPlurality}
respectively).

The SPM analyze the behavior of \textquotedblleft actors\textquotedblright ,
which can stand for formal political parties, informal factions, prominent
families, nation states, ethnic groups, social classes, industry groups,
and so on. The issues over which they struggle and/or negotiate are
represented as a continuum of choices, e.g. the degree of pro-US or
anti-US alignment, price of a cartel\textquoteright s commodity, level
of social services, location of facilities, level of troop commitment
to a war, and so on. The general ability of each actor to influence
an outcome is their \textquotedblleft capability\textquotedblright ,
which could reflect religious authority, available campaign funds,
armed soldiers, size of an ethnic voting bloc, GDP, or whatever is
relevant to the situation being analyzed. Voting is generalized from
a formal process of counting ballots to the more general concept of
exerting capability in order to influence outcomes: ``voting with
their feet'', ``voting with their guns'', expending campaign funds
to defeat a political opponent, expending strike funds to coerce an
employer, and so on. The set of actors, the issues, and their capabilities
are generally estimated by careful expert analysis before the SPM
are applicable. The actors\textquoteright{} capabilities would then
summarize the results of those analyses. Expert interpretation of
the model output is also necessary, as the true meaning of the output
depends on how the inputs were defined and estimated.

A great many studies support the idea that SPM are accurate descriptive
models of US domestic politics, especially spending issues. \cite{IPV-in-Congress}
analyzed all the recorded votes in the US Congress and concluded that
80\%-90\% can be explained by the uni-dimensional SPM. 

The SPM predict that there will be a clear order of public choice,
without cyclic preferences, which was confirmed analysis of Congressional
votes over district specific grants \cite{UnstableCD}. Several studies
\cite{SocSecGrowth,PolEconHwy} found strong evidence that SPM are
better models of legislative decisions on large scale public programs
than are other interest group models.

Nevertheless, there are well-known situations, both theoretical and
practical, in which the SPM could be improved upon. Some issues are
normative and some are descriptive; this work considers only the descriptive.
Many overall discussions of the problems exist, including \cite{RelevanceMVT,MVTinPC,CWMyth}
. Some of the issues will be discussed below in the context of explaining
the NPCE. To see how the NPCE methodology is fundamentally different
from the SPM, we examine an even older analytic construct on which
they are both based: the concept of a Condorcet winner (CW). Just
before the French Revolution, \cite{Condorcet} introduced the CW
in the context of designing group decision procedures (especially
elections) which were fair according to some unarguable \textquotedblleft gold
standard\textquotedblright . A deterministic \textquotedblleft Condorcet
election\textquotedblright{} (DCE) is the infeasible procedure of
pairing every option against every other in one-on-one contests; the
CW is that option which would win against every other option. Any
feasible choice procedure which always chooses the CW is Condorcet
consistent. 

Even in theory, not every election has a CW, though several special
kinds of elections do. In 1950, Duncan Black published a special case
of Condorcet's approach, the MVT, which assumes the following \cite{MVTheorem}:
\begin{itemize}
\item Policy options (positions advocated and outcomes experienced) can
be expressed as a single number (e.g. level of troop commitment) which
is one-dimensional, bounded, and continuous.
\item While each actor may have a different preference function, each function
must be unimodal and monotonically declining in distance from their
most-preferred option
\item In any choice between two options, each actor exerts all their capability
toward whichever is preferable to them. Because each utility curve
slopes down away from the actor's position, this means that they vote
for whichever option is closer to their own.
\item The group vote is the sum of the individual votes between options.
\item Group choices between options are deterministic.
\end{itemize}
The MVT states that under these conditions, a CW does exist and is
the median voter position, where half the total capability is on either
side. Notice that if there are gaps between the positions of different
groups, the actor holding the CW can shift its position and thus shift
the group outcome. Thus, they have the power to control the outcome,
as long as they do not move so far that some other actor becomes the
CW. While some actors are inflexibly committed to hold their position,
some actors are motivated by the desire to hold the decision power
and enjoy its benefits. For the latter, there is a strong incentive
to occupy the CW position. 

Because the CW cannot be defeated by any other proposal, it is the
predicted outcome of a group bargaining process. Any actor proposing
an option closer to the CW has a winning proposal, power-seekers always
have a motive to do so, and once the CW is proposed, nothing can defeat
it. Thus, with free introduction of proposals, the CW is the predicted
outcome of a group bargaining process. In particular, when the \textquotedblleft options\textquotedblright{}
are different candidates running for election, this means first appealing
to the median voter of their party\textquoteright s primary election,
then appealing to the median voter of the overall multiparty election.
The fact that those two medians usually differ appears to be the origin
of the popular political maxim that politicians must always \textquotedblleft run
to the center\textquotedblright{} after the primary.%
\footnote{A great deal of political commentary and analysis deals with the issues
of how to \emph{effectively} stake out a position near the median
of the primary electorate even though primary voters expect a run-to-the-middle
later, and how to \emph{effectively} stake out a position near the
median of the general electorate even after the general voters saw
the primary position. The tactics of how best to adopt positions or
to exert influence are not considered in this paper.%
}

Some of the SPM generalize this approach somewhat to a multidimensional
space, but they still require unimodal, monotonically declining preferences
on a continuous set of options. The status of multidimensional SPM
is quite complicated, largely because Generalized Mean Voter Theorem
states that the multidimensional analog of the MVT holds only under
restrictive symmetry conditions.

\subsection{Normative vs. Descriptive\label{sub:Normative-vs.-Descriptive}}

In our context, the CW is taken as purely descriptive, not normative.
While the CW may have some normative content in some contexts, it
is important to understand that it cannot be assumed that the CW represents
any sort of collegial agreement or compromise pleasing to both sides.
It might represent an explicit agreement between negotiators, or it
might represent the outcome of a group process in which no party explicitly
negotiates. 

Similarly, the ``election'' and ``bargaining'' process could be
formal, explicit, and professional, or it could be the implicit outcome
of group dynamics. We do not assume the absence of coercion, so the
relative capability of parties will greatly affect the bargains and
compromises reached. Where coercion is possible, the dynamic is similar
to an out-of-court settlement or an armed threat: If actor A is much
stronger than B, he can unilaterally create a situation of potential
conflict. In this case, the ``bargain'' to avoid a clash will likely
be a very harsh one which B would have avoided if he were stronger.
Consider the example of an armed robbery. The ``reference situation''
of open conflict is that the victim dies and the robber takes the
money. For the victim to surrender the wallet is the outcome of Nash's
model of bargaining because each is better off than the reference
situation (see \cite{NashBargaining}): the victim loses their money
but stays alive, and the robber gets the money and avoids a murder
charge. Where coercion is not possible, the reference situation is
simply the status quo in case of no agreement, or what the parties
can achieve unilaterally; this is what negotiators refer to as ``best
alternative to a negotiated agreement'' or BATNA.

As discussed in \cite{MassActionNashEql}, Nash emphasized that his
model was applicable to conscious agents with explicit foresight (the
``rational actor'' model) as well as more unconscious, evolutionary
processes (the ``mass action'' model). While bargains could be explicitly
negotiated as written agreements, they also could implicitly evolve
without communication, such as the evolution of tacit truces between
artillery on opposite sides of the trenches in WWI. An example implicit
process would be the continual adjustment of each side's behavior
and propaganda to the other side's behavior and propaganda, even if
neither side thought of itself as explicitly bargaining; see section
\eqref{eq:FC-third-party-votes}. 

A simple example was given (from the Northern perspective) in Lincoln's
Second Inaugural address:
\begin{quote}
\emph{Both parties deprecated war, but one of them would make war
rather than let the nation survive, and the other would accept war
rather than let it perish, and the war came.}
\end{quote}
This can be seen as the CW between the North and the South, where
each preferred peace to war, but each preferred war to the other's
peace. Hence, war is the CW option, even though neither party alone
would have chosen it. Thus, the issue set must be constructed to include
a wider set of options than either actor would choose alone, as it
must also include both desirable and undesirable ``compromises'',
the results of failure to formally agree (e.g. the coming or continuation
of civil war, the continuation of status quo if no legislation is
passed), and so on. In the SPM, all possible positions can be represented
by numerical coordinates, but non-spatial models raise subtle issues.

Some authors argue a desirable property of normative decision rules
is join-consistency: if group $A$ alone chooses option $X$, and
group $B$ alone chooses option $X$, then groups $A$ and $B$ together
should also choose option $X$. Some argue that the mere fact of merger
might in some cases change the context, so that join-consistency cannot
be specified as an iron-clad rule for all situations. From a descriptive
perspective, it is easy to cite examples of join-inconsistency, such
as two small, effective organizations which merge to form one larger
organization that takes ineffective actions which neither component
would choose. This is illustrated by the proverb that ``a camel is
a race horse designed by a committee'': each member might design
their own kind of race horse, but because they are different the group
can not agree on a single good design and incorporates contributions
from each member, even when they are inconsistent with each other. 

Deliberately echoing the Civil War example, each side preferred a
horse to a camel, but each preferred a camel to the other's horse.
This can happen because any complex option has a large set of possible
attributes. When considering the infinite space of designs, group
$A$'s design of a ``fast racer'' might be quite different in detail
from group $B$'s design of a ``fast racer''. The fact that both
are denoted as ``option $X$'' obscures the fact that they might
diverge in very significant ways, though they share the abstract characterization
of ``fast''. Again, North would have chosen ``peace'' and the
South would also have chosen ``peace'', but their ideas of peace
were so divergent that the CW of both together was war. From a purely
non-normative perspective, the requirement of join-consistency appears
to be violated by actual examples.%
\footnote{It could be objected that this is a false counter example, because
there were two distinct options, ``Northern-peace'' and ``Southern-peace'',
not a single option ``peace'' which was selected by both North and
South alone. By this logic, join-consistency is preserved, because
there never was a single option which each alone would select. However,
this same logic calls into question the entire rationale of join-consistency,
because real world agreements on a single option almost always mask
differences in detail of definition and implementation, so that the
two parties never choose exactly the same option down to the last
detail: there are essentially always distinct options, and the pre-conditions
of join-consistency are almost never met. Apparently, the most that
can be said is that join-consistency applies in those cases where
the differences are not important and it does not apply in those cases
where the differences do matter.%
}

The design example  raises the issue of explicit enumeration of small
issue sets versus implicit definition of infinite issue sets. More
discussion can be found in section \ref{sub:Locating-a-CW}.

\section{Overview of NPCE}

The \textquotedbl{}non-spatial probabilistic Condorcet election\textquotedbl{}
(NPCE) methodology starts with the same concept of a Condorcet Election,
but diverges even before the MVT. The NPCE methodology treats the
group choices between options as probabilistic, rather than deterministic.
It generalizes the deterministic Condorcet approach to handle a more
general set of cases at the cost of providing less specific predictions
and requiring somewhat more computational resources. Given modern
computing infrastructure, e.g. on-demand provisioning via cloud computing,
this is not a serious limitation.

The NPCE methodology is open to specialization for problems with unique
structure, analogous to the situation with linear programming. Linear
programming (LP) is a very general and powerful methodology, but many
problems have special structures that are utilized by specialized
algorithms to compute the same result more efficiently. Classic LP
examples include transshipment algorithms which assume a network of
balanced, positive flows. It is well-known that deterministic CW exist
and can be efficiently computed for certain network location problems
\cite{Variations,VotingandPlanning}. Some macroeconomic problems
also have CW, under restrictive assumptions about the actors' utility
functions (e.g. all identical Cobb-Douglas); examples are discussed
in \cite{IncomeIneq-PubGoods}. More directly, there is a large literature
on exploiting matrix structure to compute various kinds of eignevectors,
which is quite similar to the core NPCE problem of computing a limiting
distribution over possible CW; we fully expect that analogous special
structures can be utilized for NPCE.

The new NPCE methodology extends the state of the art in several ways:
\begin{itemize}
\item Non-deterministic: The methodology is designed to identify likely
outcomes and interactions, as well as estimate the associated risk,
opportunities, and variability.
\item Discrete Options: Support the analysis of not only continuous gradations
of policy but also discrete, combinatorial choices
\item The Value of Outcomes: Extend the state of the art to look at the
value to actors of real-world outcomes generated by policies, not
just differences in the policies.
\item Differing perceptions of policy space: The NPCE methodology does not
require that the parties to a dispute all have the same characterization
of how choices are ordered or placed.
\end{itemize}

\subsection{Deterministic vs. Non-deterministic\label{sub:Deterministic-vs-Nondeterministic}}

The SPM appear to be built on the assumption that political processes
are largely deterministic. In a legislative vote where one position
gets 11 votes and the other get 10, then the former is the unambiguous
winner. The published analytical models do not appear to include any
provision for uncertain outcomes and appear to use expected-value
computations throughout, ultimately producing a single point prediction
of a single forecast outcome, given particular input parameters. Much
of the literature on CW and probability appears to focus upon the
probability that a CW will result from a deterministic CE, for various
models of stochastic voter preferences. That is, the CW is considered
as a deterministic function with stochastic parameters.

While deterministic outcomes seem quite realistic when analyzing formal
elections with ballots, it becomes questionable when comparing the
generalized ``voting'' mentioned earlier. For example, if two equally
strong interest groups try to influence legislators to get their preferred
outcome, then either one is equally likely to prevail. This can be
taken as the operational definition of \textquotedblleft equal capability\textquotedblright ,
as is often done in systems to rank the capability of chess players,
sports teams, and so on. Thus, expert opinion or prior statistics
on relative odds can be used to derive relative capabilities via paired
comparison techniques. When they are slightly mismatched, the odds
can be expected to shift slightly, rather than snapping to a deterministic
all-or-nothing outcome. This suggests that the CW could be considered
as a stochastic function with deterministic parameters.

The simplest model is that the probability of success in each one-on-one
contest depends on the relative capabilities of the actors involved:

\begin{equation}
p\left[i\succ j\right]=\frac{c_{i}}{c_{i}+c_{j}}\label{eq:Basic-bilateral-victory-prob}
\end{equation}

Zermelo's original chess ranking function \cite{Zermelo}, the Bradley-Terry
model \cite{BradleyTerry}, the Glicko model \cite{Glicko} for sports
teams and many others \cite{BayesianOnlineRanking} are examples.
While many alternative variations are discussed in the literature,
most are transformations of equation \eqref{eq:Basic-bilateral-victory-prob}.
However, the core of the NPCE methodology is the technique of using
probabilistic contests; equation \eqref{eq:Basic-bilateral-victory-prob}
is a sub-model which can be varied based on the problem at hand.

Unlike methodologies based on DCE and the MVT, the NPCE methodology
is designed from the foundations to represent this stochastic aspect
of political/military/economic struggles. The standard \textquotedblleft forecast\textquotedblright{}
is a probability distribution over the possible outcomes. A disadvantage
of the NPCE is that while the MVT provides the strong prediction of
a single outcome, the NPCE provides a weaker prediction of probabilities.%
\footnote{When the CW exists, it is usually ranked as the most probable outcome.
When the CW defeats other options by wide margins, it is much more
likely than other options. When the CW just barely defeats other options,
it may be slightly more or less likely than the alternatives.%
}

Adopting a probabilistic perspective is a fundamental change from
the DCE and MVT and work built upon them. In particular, the strict
problem structure necessary to make the strong prediction of a single
outcome is no longer necessary. This one change opens the door to
a broad range of new problems areas including (but not limited to)
multidimensional choices, discrete choices, and domain-specific utility
models. This breadth comes at a price. First, less precise conclusions
can be drawn. Second, with only statistical predictions, validation
is more challenging. Third, models of dynamics and bargaining tailored
to DCE and MVT, in SPM, are no longer applicable.

\subsubsection{Strategy Optimization}

Deterministic SPM often exhibit the following properties. Because
a small shift in the CW actor shifts the predicted outcome, while
small shifts in all other actors has no effect, efforts to use SPM
to plan strategies often focus on actors at the median position, especially
when they are powerful. With discrete options, strategy optimization
with deterministic models has the difficulty that the response surface
(predicted outcome as a function of the actor capabilities) is discontinuous,
as the prediction is either one discrete outcome or another: the response
is everywhere perfectly flat except where it is discontinuous. 

With NPCE, the prediction is not a single outcome but a distribution
over outcomes. Even for the case of discrete options, the distribution
is a continuous function of the input parameters, so the expected
value of the result is a continuous function of the input parameters.
The availability of continuous slope information makes it possible
to bring many powerful optimization techniques to bear in order to
suggest novel strategies.

In the case of one-dimensional models, the NPCE can be used to evaluate
strategies that do not necessarily focus immediately on the CW actor.
While influencing the first-tier actors is always an option to be
analyzed, the NPCE makes it easier to analyze the potential for shifting
the odds in one's favor by influencing second-tier actors.

The NPCE methodology also facilitates analyzing the robustness of
outcomes and strategies, for both continuous and discrete issue sets.
For example, suppose an analyst is looking for strategies to influence
a group so as to select option X. The predicted distribution for strategy
A might indicate option X was the most likely, but only by a slim
margin. With a different strategy B, option X might still be the most
likely, but by a large margin. While a deterministic analysis might
simply indicate option X in both cases, the NPCE will provide the
additional information that strategy A is more risky, in the sense
of having higher RMS deviation from one's goal.

\subsection{Continuous vs. Discrete Options}

The SPM represent multiple political choices as points in a continuous,
multidimensional space, where each dimension represents an independent
sub-issue. Each point defines an overall policy which an actor could
advocate and/or a policy compromise among several actors. A large
class of problems which do not fit the SPM paradigm are subset selection
problems (SSP). We will consider two: interest groups selecting parties
to form a government, and managers choosing projects to form a research
program.

The composition of a government coalition is discrete, not continuous:
each party is either in or out. Similarly, the assignments of cabinet
seats to coalition members are inherently discrete: it is impossible
to assign 2.3 cabinet seats to a party. Thus, neither of these problems
can be represented well in SPM. The problem cna be represented as
multistage, because the choice of government in the first stage strongly
influences the policy outcomes in the second stage.

Because each party has a slate of policy preferences, a choice between
one ruling party and another might have implications for several issues
at once: free markets vs. central planning, extensive or minimal rights
for women, orientation toward America or Russia, and so on. Because
there are generally more policy issues in play than there are parties,
choosing which party to back (or which party to get a cabinet seat)
inevitably involves tradeoffs between those issues. Notice that this
is a completely different mechanism to address policy tradeoffs than
is used in multidimensional SPM.

Once in power, the parties may cooperate via \textquotedblleft logrolling\textquotedblright{}
pairs of issues where each does a favor for the other. Finally, all
these interactions are anticipated in the struggles over how to form
a government, as every actor is trying to shape the final outcome.
The NPCE methodology can be used by analysts to examine how the actors
involved might address the discrete choices in forming a government,
in light of their understanding of the multistage nature of the process.

There are fundamental reasons why the discrete structure of a government
cannot be easily summarized into a single number, greatly limiting
the ability of SPM to address these tradeoffs. When there are a half-dozen
parties competing to get into the government, then each alternative
outcomes is simply a subset specifying which get in and which are
left out. This can be represented by a simple list of 1's and 0's,
which\emph{ }is more like the evenly spaced corners of a unit hypercube
than a continuous line segment.\emph{ }Any scheme to reduce it to
a simple continuum (e.g. Hamming distance from a given corner) inevitably
maps significantly different points onto each other, thus obscuring
the difference between alternatives. With $m$ parties, there are
$2^{m}$ options for which subset is \textquotedblleft in\textquotedblright{}
and which is \textquotedblleft out\textquotedblright , with the widest
diversity of options at the level of 50\% in and 50\% out. An analogy
would be to categorize every point on the globe by its distance from
the Entebbe airport and then placing them on a line: places as dissimilar
as the North Pole, Vietnam, the Indian Ocean, the South Pole, and
the North Atlantic would all collapse to the point representing \textquotedblleft halfway
around the world\textquotedblright . 

While policy options can be mapped to {[}0,1{]} vectors and thus to
the corners a hypercube, the utilities common in SSP are poorly represented
by a weighted Euclidean distance; an example is discussed below. A
generalization of the SSP is to determine a matching. Given a dozen
factions and six cabinet seats, each possible matching can be represented
by a six-element list of which factions get to control each seat.
With $n$ seats and $m$ factions, there are $n^{m}$ possible ways
to match them up. Because neither seats nor factions are naturally
ordered, (unlike the 0/1 alternatives for whether a party is in the
coalition or not) there is no obvious way to place all the matchings
on the corners a hypercube. Again, there is (in general) no way to
reduce a matching to a single number without potentially losing critical
information. For example, a large pro-US faction might be forced to
cooperate with a small but very anti-US faction by granting them one
out of five seats in the cabinet, so that every possible cabinet is
80\% pro-US and 20\% anti-US. However, it is important whether that
seat controls farm policy or the entire military: we would like to
disaggregate the single 80\% number into a matching of factions with
seats.

\subsection{Policies vs. Outcomes}

SPM postulate that the attractiveness of a policy option to a decision
maker depends on the weighted Euclidean distance between that policy
option and the decision maker\textquoteright s preferred option: their
own option is best, and desirability declines as \textquotedblleft distance
in policy space\textquotedblright{} increases. Technically, their
utility function is unimodal and monotonically declining with distance.
Thus, to fit a problem into the SPM framework, one must define a single
continuous space (and weighted distance measures on it), so that every
actor\textquoteright s preference function becomes unimodal and monotonically
declining. As described below, there are many important cases when
this is not possible even with continuous options.

\subsubsection{Multiple Peaks}

It frequently turns out that the desirability of a policy choice is
not single-peaked in any distance measure. As we will demonstrate,
this can be a problem even for unidimensional issues.

One classic example is a policy choice over the level of troops to
commit to a war. One actor's best option might be to commit a high
number of troops in order to win decisively, next-best to commit a
low number as non-provocative observers, and least desirable to commit
a middle amount: enough to suffer large casualties but not enough
for a significant chance of victory. For this actor, the preference
ordering of policies, based on military outcome, be $H>L>M$. When
this actor\textquoteright s utility is graphed versus number of troops,
there are peaks at both extremes and a dip in the middle (the proverbial
``worst of both worlds''). 

The case of public bads has been analyzed in \cite{Three-Essays}.
This situation is characterized by a unidimensional scale, where each
actor's utility function has a single triangular dip, with perfectly
flat indifference regions on either side. The situation of multiple
dips or peaks in each utility function, or having multiple dimensions
or discrete options, is not analyzed.

Because actors have different utility functions, each actor might
have different number of dips and peaks, located in different places.
The SPM are inherently unable to represent actor\textquoteright s
preferences in such situations where benefit has multiple peaks. The
fundamental problem is that actors often care about outcomes that
are non-monotonic in policy.

\subsubsection{Multiple Orders}

Frequently, actors will value options by different criteria, because
they value the outcomes by different criteria. This is just one of
several reasons why it is generally impossible to map discrete options
into a single distance measure for all actors: while Hamming distance
might be appropriate for one  actor\textquoteright s valuation of
the consequences, it is not therefore automatically appropriate for
another actor who values the consequences differently. In the example
of troop commitments, one actor, $A$, might order the desirability
of alternative by the military outcome, giving $H>L>M$ discussed
earlier. Another actor, $B$, might be concerned with cost, giving
the $L>M>H$ ordering. Of course, for any given actor, the alternative
policies could be sorted according to how good the outcome was for
that actor (with some overlaps). It is simple to enumerate all six
orders of ${L,M,H}$ to verify that none make $A$'s utility function
unimodal and $B$'s unimodal at the same time.%
\footnote{This situation is superficially similar to Arrow's Impossibility Theorem
\cite{AIT}. However, Arrow's theorem applies when each individual's
preferences are defined by a total order without numeric weights,
so it is not applicable to situations where choices have numerical
weights, such as range voting or the generalized voting considered
in this work.%
}

Because actors can have arbitrary preferences over outcomes, there
is in no way to define any distance measure which gives the correct
preference ordering for all actors when combined with a unimodal,
monotonically declining utility measure. Even in the uni-dimensional
continuous case, there is not always an embedding of the options so
that they are placed in one order for one actor, while appearing to
be in a completely different order for another actor; the troop deployment
problem provides an example. The NPCE methodology completely avoids
this problem by working directly with actor\textquoteright s individual
utility of outcomes, without trying to collapse them all into a single
measure of distance in a common policy space.

\subsubsection{Domain Specific Utilities}

The SPM treat options basically as points in continuous space, where
some measure of distance determines utility. Even when actors embed
policy options in space differently, the basic spatial concept applies.
However, there are many cases where the spatial analogy is quite strained.
A simple example is the SSP of which research projects to fund and
which not to fund. Each option can be represented as a discrete choice
of which subset of projects to fund. Relative to a set of research
goals, the utility of the top-level discrete choices depends not on
a measure of \textquotedblleft distance in policy space\textquotedblright{}
but on an estimate of the real-world consequences. Because of the
potentially complex precedence and complementarity relations between
projects, utility can be highly non-monotonic in any total or partial
order (rather like a combinatorial auction). For selection of defense
projects, complex non-monotonic considerations of tactics, available
industrial base, strategy, and geography are necessary: detailed modeling
and analysis are often required before the summary assessments of
utility can be used as actors' estimates of the utility.

Of course, it is possible that a more bureaucratically minded actor
might evaluate options purely by how many of their agency's favored
programs are included and how many of rival agencies\textquoteright{}
programs are excluded. Because the NPCE methodology does not assume
actors have the same utility function, including both research oriented
and bureaucratically oriented actors presents no difficulty.

The NPCE methodology represents this situation by using sub-models
which mimic the process by which actors anticipate the consequences
of their choices, and value their choices according to their expected
value of the consequences. In the case of funding research tasks,
one plausible sub-model would be a probabilistic PERT chart for the
tasks. It not assumed that each actor has the same model, that any
are objectively correct, or even that they agree on what the consequences
are likely to be. The model of their expectations is termed a domain
specific utility model (DSUM).

As mentioned earlier, the choice of what government to form (or which
cabinet positions to assign to which coalition members) can be modeled
as turning on an estimate of the various policies which might result
from that government. The example DSUM considered in this case assigns
values to each top-level choice of government by invoking sub-models
for each issue; each sub-model is another copy of the NPCE for the
corresponding one-dimensional issue, parametrized for that particular
government\textquoteright s composition. The RMS deviations from that
actor's ideal result on each issue are separately estimated (i.e.
no logrolling across issues by different members of the government),
then combined with the actor's salience for each issue to derive a
final utility for that government. Thus, the DSUM mimics the actor\textquoteright s
foresight as to the likely range of outcomes from their choice of
government.

Another example is the design of tax and subsidy policies or economic
stimulus packages in a multiregional country: there are multiple groups
affected by each policy choice, groups are often affected in different
ways, regions are affected in different ways, and effects can easily
be non-monotonic in the size or targeting of a policy. Political actors
are likely to weigh the benefits to their own constituency heavily,
while weighting the effect on others\textquoteright{} constituencies
quite differently. Similar considerations apply to multinational models.

\section{Details of NPCE}

Rather like linear programming, the core NPCE methodology is simple
and abstract enough to be stated in just a few equations: \eqref{eq:Basic-Coalition-Strength},
\eqref{eq:Coalition-victory-prob}, and \ref{eq:Complex-Transition-Eigenprobs}.
This section is devoted to examining the implications and usage of
the equations, as well as presenting various options for some of the
sub-models which support the equations. Like linear programming, there
is an ``art and science'' of building a model within the framework,
and another science to configuring and solving the models efficiently.

\subsection{Model Variations\label{sub:Model-Variations}}

There are many choices at each stage of the model-building process.
Basic structure of the issue set, domain specific utility functions,
voting rules, third-party commitment, and search strategy are just
some of those available. Almost all combinations yield a model in
which no deterministic CE exists, necessitating the use of probabilistic
voting models as explained in section \eqref{sec:Problem-and-SOA}.
The basic NPCE methodology has been implemented in several prototypes
whose main purpose is to explore the effect of varying those assumptions
in both continuous and discrete issue sets, so that in each problem
the important assumptions can be identified for further study. With
its ``Chinese menu'' of options, more model structures can be created
than can be discussed here: most variants are not discussed in the
literature, some new variants are likely to be useful, but many will
not.

For each of several sub-models, each prototype supports several sub-models
(e.g. voting rules from equations \eqref{eq:Binary_Voting_Between_Alternatives},
\eqref{eq:Proportional_Voting_Between_Alternatives}, \eqref{eq:Cubic_Voting_Between_Alternatives}
and more), but we will generally limit discussion in this paper to
sub-models whose behavior can be described in fairly simple equations
(e.g. binary or proportional voting). Thus, many of the equations
which follow describe the \emph{behavior} of the generic numeric methods
when applied with particular sub-models (e.g. voting rule, third party
commitment, etc.) but do not appear in the software: they only describe
the behavior for that particular set of modeling options. The software
uses only the top-level equations and generic numeric methods, which
can be applied to a wider range of options than admit simple analytic
solutions.

\subsection{Issue sets}

The set of possible positions on an issue define the issue set. It
may be explicitly given, such as the set of all integer values from
0\% to 100\%, or it could be an implicit definition of an enormous
combinatorial set, such as the set of possible organizational charts.

For NPCE, no more structure is required. The key steps in defining
an issue set for SPM are the following:
\begin{itemize}
\item Define a distance metric, $d$, on the set of alternatives, $\{\theta_{i}\}$.
The most common way to do this is to associate one or a few numerical
coordinates with each option, then use a weighted Euclidean distance
metric on those coordinates. Any distance metric must obey the standard
definition:

\begin{itemize}
\item Non-negative: $d(\theta_{i},\theta_{j})\geq0$
\item Identity of indiscernibles : $d(\theta_{i},\theta_{j})=0\Leftrightarrow\theta_{i}=\theta_{j}$
\item Symmetry: $d(\theta_{i},\theta_{j})=d(\theta_{j},\theta_{i})$
\item Triangle inequality: $d(\theta_{i},\theta_{k})\leq d(\theta_{i},\theta_{j})+d(\theta_{j},\theta_{k})$. 
\end{itemize}
\item Define an indexed family of utility functions which are declining
in distance: $d(\theta_{i},\theta_{j})<d(\theta_{i},\theta_{k})\Leftrightarrow U_{i}(\theta_{j})>U_{i}(\theta_{k})$
\end{itemize}
Notice that, taken together, these conditions imply that each actor
ranks their own position higher than any other distinct position:
$\forall\theta_{i}\neq\theta_{j}\: U_{i}(\theta_{i})>U_{i}(\theta_{j})$

As discussed earlier, it is not always possible to construct the $(d,U_{i})$
required for the SPM.

\subsection{Voting}

Given a choice between two alternatives, $\alpha:\beta$, the effort
an actor will exert to promote one or the other is called his ``vote'';
it is determined by his ``capability'', $c_{i}$ and the difference
in utility to him of the two alternatives.%
\footnote{Note that the models in \cite{ECDecisionMaking,ForecastingPolEvents,ReportCFP}
do not fit this model, because they assign utility not to the positions
of actors but to changes in those positions. We will denote the particular
formula to calculate utility to actor $i$ of a change in state from
$S_{1}$ to $S_{2}$ as $U_{i}(S_{1},S_{2})$. It can not be represented
as the difference between utilities of individual states, because
the utility for no change at all is assigned a positive value; see
\cite{ForecastingPolEvents} page 52. While $U_{i}(S_{1},S_{1})>0$
for every state $S_{1}$, there is no function $u$ so that $u(S_{1})-u(S_{1})>0$.
Therefore, the work described in this paper does not apply to those
models or to models based upon them.%
} As mentioned earlier, ``voting'' is generalized to mean the exertion
of capability in order to affect outcomes. Positive favors option
$\alpha$, while negative favors option $\beta$. Voting can be over
many kinds of alternatives: positions currently held by actors, alliances
with other actors, bargains struck between actors to define new positions,
and more. We follow the von Neumann convention of bounding utility
by $0\leq U_{i}(\alpha)\leq1$.

\subsubsection{Voting Rules\label{sub:Voting-Rules}}

The binary voting rule is as follows:
\begin{equation}
v_{i}(\alpha:\beta)=\begin{cases}
+c_{i} & \quad\mathrm{if}\: U_{i}(\alpha)>U_{i}(\beta)\\
-c_{i} & \quad\mathrm{if}\: U_{i}(\alpha)<U_{i}(\beta)\\
0 & \quad\mathrm{if}\: U_{i}(\alpha)=U_{i}(\beta)
\end{cases}\label{eq:Binary_Voting_Between_Alternatives}
\end{equation}

This all-or-nothing behavior is a good model of casting yes/no votes
(weighted or not), but in many cases of exerting informal influence,
a more nuanced exertion of effort is observed. One simple model of
nuanced voting is that actors exert effort in strict proportion to
what is at stake for them between the two alternatives:
\begin{equation}
v_{i}(\alpha:\beta)=c_{i}[U_{i}(\alpha)-U_{i}(\beta)]\label{eq:Proportional_Voting_Between_Alternatives}
\end{equation}

For example, environmental interest groups might spend only a small
amount of effort on unimportant issues, while expending a great deal
of time, money, and attention on issues they view as critical. As
derived in \cite{Lodestone,GeneralIssueSpaces} and published in \cite{FutureOfIran},
proportional voting leads to CW under extremely general conditions.
The proof is quite simple:

\begin{equation}
\begin{array}{ccc}
V\left(\alpha:\beta\right) & = & \sum_{i}v_{i}(\alpha:\beta)\\
 & = & \sum_{i}c_{i}[U_{i}(\alpha)-U_{i}(\beta)]\\
 & = & \sum_{i}c_{i}U_{i}(\alpha)-\sum_{i}c_{i}U_{i}(\beta)\\
 & = & \omega\left(\alpha\right)-\omega\left(\beta\right)
\end{array}\label{eq:CPT-proof}
\end{equation}

where

\begin{equation}
\omega\left(x\right)=\sum_{i}c_{i}U_{i}(x)\label{eq:Benthamite-Social-Utility}
\end{equation}

Similar to Black's original proof of the MVT, actors seeking office
always have a motive to propose options with higher $\omega$ values,
leading the group to the ``Central Position'' as the maximum; we
call this the Central Position Theorem (CPT). The difficulty of maximizing
$\omega$ can range from very easy to very difficult, depending on
the properties of the utility functions and set of options. A special
case is analyzed in Corollary 4.4 of \cite{Coughlin-PVT} assuming
a probabilistic Luce model of weighted binary voting, over a continuous,
convex, compact set of options with a particular exponential form
for the probabilities. 

With the von-Neumann scaled utilities of this paper, the resulting probability of the i-th actor 
voting for option $\alpha$ over $\beta$ is simply the following:

\begin{equation}
p_{i} (\alpha:\beta) ={{ 1+U_{i}(\alpha)-U_{i}(\beta) } \over 2 }
\label{eq:Prop-vote}
\end{equation}

When the actual votes between options are set
equal to their expected values under these assumptions of  \cite{Coughlin-PVT}, then the position
maximizing equation \eqref{eq:Benthamite-Social-Utility} is the voting
equilibrium.

While proportional voting is a common modeling choice, there are well-known
cases where the proportional rule does not appear to hold. There is
a well-known lobbying rule to ``focus benefits and diffuse costs''
which suggests that equation \eqref{eq:Proportional_Voting_Between_Alternatives}
does not hold in this case. Suppose the US Congress creates a program
which draws 1.5 billion dollars from general tax revenue to confer
benefits on a constituency of ten thousand people. The benefit to
the constituents is \$150,000 to each of 10,000 people. If there are
effectively 150 million taxpayers, then the cost is only about \$10
each. Assuming for now that utility is proportional to dollars and
that each affected individual has equal capability to exert influence,
the strict linearity of equation \eqref{eq:Proportional_Voting_Between_Alternatives}
implies that the influence of the few but highly motivated supporters
would be exactly neutralized by the net influence of the numerous
but weakly motivated opponents. However, it is well known that such
diffuse costs are largely ignored by those effected, leading to the
lobbying advice. To model this situation, the cubic rule is designed
to have a shallow slope for small changes, but steep slope for large
ones. This reflects the lobbying rule so that voters harmed are in
the shallow region while voters helped are in the steep region.
\begin{equation}
v_{i}(\alpha:\beta)=c_{i}[U_{i}(\alpha)-U_{i}(\beta)]^{3}\label{eq:Cubic_Voting_Between_Alternatives}
\end{equation}

One can imagine many other alternative voting rules, such as linear
mixture of cubic, proportional, and cubic. Most such rules have not
been extensively studied; this may be because they are not algebraically
tractable, do not reliably produce CW with DCE in common cases, or
both. Though it seems to reflect the empirical experience of lobbyists,
this author has found no discussion of the cubic voting rule, or its
CE/CW properties, in the literature.

Regardless of the individual voting rule, the net vote of the group
between the alternatives is just the sum of the individual votes.
Note that inter-subjective comparisons of utility is never done; only
capability to exert power is compared. This is why it is important
that the changes in utilities be in a fixed range (e.g. $[-1,+1]$
models with utility on a $[0,1]$ scale): the limits of actions by
actors should be determined by their relative capability to exert
power, not by the scale on which their utilities happen to be measured.
\begin{equation}
V(\alpha:\beta)=\sum_{i}v_{i}(\alpha:\beta)\label{eq:Group_Voting_Between_Alternatives}
\end{equation}

If this quantity is positive, then we say that the group prefers $\alpha$
over $\beta$, though weaker players may strongly disagree. In this
sense, the stronger actors can force through outcomes against the
wishes of weaker actors. This is normal and expected in democratic
elections: if one interest group (aka ``actor'') gets 70\% of the
vote, then the majority wins, even if the remaining 30\% still would
have preferred the other candidate. On the other hand, the CW might
not be very desirable to any actors, even though it is the best the
group can decide upon.

As mentioned above, the strong CW is defined to be that option $\alpha$
which defeats all others:

\begin{equation}
\forall\beta\neq\alpha\; V(\alpha:\beta)>0
\end{equation}

The weak CW loses to none: 
\begin{equation}
\forall\beta\; V(\alpha:\beta)\geq0
\end{equation}

Because this allows the possibility of very weak ``winners'' (e.g.
all options are tied), some authors extend this to require that a
weak CW is never defeated and defeats at least one:
\begin{equation}
\forall\beta\; V(\alpha:\beta)\geq0\;\wedge\;\exists\gamma\; V(\alpha:\gamma)>0
\end{equation}
 This allows only non-trivial ties, such as A defeats C and ties B,
while B defeats C and ties A, where both A and B are weak CW. Further
refinements of the CW criteria exist but are not necessary for our
purposes.

As discussed in section \eqref{sub:Deterministic-vs-Nondeterministic},
discrete jumps between options are  appropriate to actual legislative
votes, while smoothly changing probabilities are more appropriate
to generalized voting. An advantage of the NPCE over DCE is that NPCE
distinguishes between a strong CW which just barely defeats the other
options (marginal) and a strong CW which soundly defeats all others
(robust). In the marginal case, the CW is estimated to be just barely
more likely than the second-most likely, while in the robust case
the CW is estimated to be much more likely than the second-most. For
our purposes, the distinction between marginal and robust CW is more
important than the distinction between strong and weak CW. Consider
the case that option A barely defeats B and soundly defeats C, while
B barely loses to A and soundly defeats C. In the DCE methodology,
option A is ranked as the strong CW no matter how small the margin
of victory, even when they are insignificant round-off errors, until
the discrete transition from ``strong'' to ``weak'' when the tiny
margin becomes exactly zero. When there is even a tiny margin of defeat
(even round-off error), it is classified as a Condorcet loser. With
NPCE methodology, there are no discrete jumps between options as the
strength of their support varies, just smoothly declining probability
from very high odds to nearly even.

When the context is clear, we often refer to the current positions
of particular actors, $\theta_{i}$, by just the index i . The effort
which actor $k$ will exert to support option $\theta_{i}$ over option
$\theta_{j}$ is $v_{k}(i:j)$. We do not consider framing effects,
so swapping $i$ and $j$ flips the sign: 
\begin{equation}
v_{k}(i:j)+v_{k}(j:i)=0\label{eq:Symmetric-individual-votes}
\end{equation}
We will consider different models of third-party voting in section
\ref{sub:Third-Party-Support}; for now we simply use $v_{k}(i:j)$.
The set of $k$ with $v_{k}(i:j)>0$ are the coalition supporting
$i$ against $j$. While each model of the third-party choice gives
a different estimate of $v_{k}\left(i:j\right)$, the total capability
of $i$'s coalition against $j$ is always the sum of votes for $i$
against $j$:

\begin{equation}
C_{i:j}=\sum_{v_{k}(i:j)>0}\: v_{k}\left(i:j\right)\label{eq:Basic-Coalition-Strength}
\end{equation}

Equations \eqref{eq:Symmetric-individual-votes} and \eqref{eq:Basic-Coalition-Strength}
imply that
\begin{equation}
C_{j:i}=\sum_{v_{k}(i:j)<0}\:|v_{k}\left(i:j\right)|
\end{equation}

Note that this includes the bilateral contributions of $i$ and $j$
themselves, which may be large or small depending on how high are
the stakes. By the symmetry of equation \eqref{eq:Symmetric-individual-votes},

\begin{equation}
C_{j:i}=\sum_{v_{k}(j:i)>0}\: v_{k}\left(j:i\right)
\end{equation}

Because every non-zero term in the sum for $V(i:j)$ appears in either
the sum for $C_{i:j}$ or $C_{j:i}$, we have 

\begin{equation}
V\left(i:j\right)=C_{i:j}-C_{j:i}
\end{equation}

In a political election, or a legislative vote, the winner is uniquely
determined by the sign of the difference in capabilities, e.g. 11
votes to 10. When the concept of exerting influence is extended from
legislative votes to more general contests of capability, then the
outcome is not so clear-cut. For example, if one legislative pressure
group has \$11 million to spend on lobbying efforts and the other
has \$10 million, then \emph{ceterus paribus} the probability of the
former prevailing in a legislative vote should be only slightly better
than 50\%. The probability that $i$ will defeat $j$ depends on the
relative capability of the coalition supporting each, via the simplest
generalization of equation \eqref{eq:Basic-bilateral-victory-prob}:

\begin{equation}
P\left[i\succ j\right]=\frac{C_{i:j}}{C_{i:j}+C_{j:i}}=P_{ij}\label{eq:Coalition-victory-prob}
\end{equation}

To avoid division by zero errors and give $50:50$ odds in the case
of exact ties, a small positive constant is added to each $C$ before
applying equation \eqref{eq:Coalition-victory-prob}. A reasonable
heuristic would be to make it a tiny fraction of the root mean square
of the $C_{i:j}$ values. This also has the useful result that $P_{ii}=\frac{1}{2}$,
which means that the purely formal contest $i:i$ has 50\% probability
of each side winning, or 100\% chance that i will remain if no other
actor challenges it.

Note that $P_{ij}+P_{ji}=1$ and that $V(i:j)\geq0$ if and only if
$P_{ij}\geq\frac{1}{2}$.

\subsection{LimitingDistribution}

The MVT and derived models use the heuristic of repeated deterministic
contests in Condorcet Election to demonstrate that group decision
making should converge to the median voter. The NPCE methodology used
the heuristic of a Markov process where repeated probabilistic contests
yield a limiting distribution. As mentioned earlier, this is not the
only Markov model of probabilistic voting. Two examples of using NPCE
combined with a Markov model to estimate a limiting probability distribution
are \cite{VotingCycles} and Zapal's PhD thesis, \cite{Zapal2012};
several seminar presentations extracted from the thesis explore \textquotedbl{}simple
Markovian equilibria in dynamic spatial legislative bargaining\textquotedbl{}.

The deterministic CW is defined by the heuristic of a full $n^{2}$
deterministic Condorcet election. The limiting distribution of a probabilistic
Condorcet election can be computed by a heuristic of ``fictitious
play''. Suppose $p_{i}^{t}$ is the probability of $i$ at turn $t$.
Then there are three ways that $i$ could be chosen at turn $t+1$:
\begin{itemize}
\item Option i was chosen at time t, challenged by option j, and i defeated
j
\item Option i was chosen at time t, and not challenged.
\item Option k was chosen at time t, challenged by option i, and i defeated
k
\end{itemize}
As mentioned above, the situation of $i$ not being challenged can
be formally represented as $i$ challenging $i$, which results with
certainty in $i$. We denote the probability of a challenge from $j$
to $i$ as $c_{ji}$, so the three conditions can be combined to give
the Markov transition probabilities as follows:
\begin{equation}
\begin{array}{rcl}
p_{i}^{t+1} & = & p_{i}^{t}\left(\sum_{j\neq i}\: c_{ji}P_{ij}\right)\\
 &  & +p_{i}^{t}\left(1-\sum_{j\neq i}\: c_{ji}\right)\\
 &  & +c_{ik}\left(\sum_{k\neq i}\: P_{ik}p_{k}^{t}\right)
\end{array}\label{eq:Markov-Trans-Probs}
\end{equation}

Because $P_{ij}+P_{ji}=1$, the assumption that $c_{ji}=1/n$ yields
the following:
\begin{equation}
p_{i}^{t+1}=\frac{1}{n}\sum_{j=1}^{n}P_{ij}\left(p_{i}^{t}+p_{j}^{t}\right)\label{eq:Complex-Transition-Eigenprobs}
\end{equation}

This can be expressed more concisely as the following vector equation:
\begin{equation}
p^{t+1}=T\left(P,p^{t}\right)
\end{equation}

Equation \eqref{eq:Complex-Transition-Eigenprobs} implies that each
term of $p_{i}^{t}$ appears in $\sum_{i}p_{i}^{t+1}$ with the coefficient
$\frac{1}{n}\sum_{j=1}^{n}\left(P_{ij}+P_{ji}\right)$, which is 1.
This implies that 
\begin{equation}
\sum_{i}p_{i}^{t}=1\;\Rightarrow\;\sum_{i}p_{i}^{t+1}=1
\end{equation}

as required.

Equation \eqref{eq:Complex-Transition-Eigenprobs} is essentially
an eigenvector problem, so we can use similar solution techniques.
One simple, general method is the iterative Newton procedure: 
\begin{equation}
p^{t+1}=\frac{p^{t}+T\left(P,p^{t}\right)}{2}
\end{equation}

Starting from a uniform $p_{i}^{0}=\frac{1}{n}$ distribution, it
efficiently converges on the limiting distribution of the Markov process,
$p=p^{\infty}$:
\begin{equation}
p_{i}^{\infty}=\frac{1}{n}\sum_{j=1}^{n}P_{ij}\left(p_{i}^{\infty}+p_{j}^{\infty}\right)\label{eq:Stationary-Eigenprobs}
\end{equation}

A little algebra shows that for the two-options case, we get exactly
the intuitive expected result:

\begin{equation}
\begin{array}{rcl}
p_{1} & = & \frac{C_{1:2}}{C_{1:2}+C_{2:1}}\\
p_{2} & = & \frac{C_{2:1}}{C_{1:2}+C_{2:1}}
\end{array}\label{eq:Markov-Limit-with-Two}
\end{equation}

With more than two options, there is no simple closed form for the
limiting distribution; the following numerical example is purely for
reference. For illustration, we have assumed that $c_{ji}=1/5$ though
other assumptions could reasonably be made. The capabilities $C_{ij}$
were generated uniformly from {[}0,1{]}, those on the middle row were
raised by 0.5 to make it the CW, and equation \eqref{eq:Coalition-victory-prob}
was used to compute each probability.

\begin{equation}
P=\begin{bmatrix}0.5000 & 0.4192 & 0.1814 & 0.8272 & 0.5211\\
0.5808 & 0.5000 & 0.3326 & 0.7129 & 0.1856\\
0.8186 & 0.6674 & 0.5000 & 0.7674 & 0.5043\\
0.1728 & 0.2871 & 0.2326 & 0.5000 & 0.1777\\
0.4789 & 0.8144 & 0.4957 & 0.8223 & 0.5000
\end{bmatrix}
\end{equation}

\begin{equation}
p=\begin{bmatrix}0.1597\\
0.1400\\
0.3401\\
0.0638\\
0.2964
\end{bmatrix}
\end{equation}

\subsection{Utility of a Contest\label{sub:Utility-of-a-Contest}}

When advantageous, any actor $i$ could initiate a one-on-one contest
in order to get actor $j$ to adopt their position. Even if the contest
does not actually occur, the possibility can be anticipated by both
parties as they consider what bargain to reach. If it were a strictly
bilateral contest, where both sides exerted their full capability,
$c_{i}$, we would expect the probability of  winning to be determined
by their relative capabilities, similar to equation \eqref{eq:Basic-bilateral-victory-prob}.
However, it is unrealistic to model every conflict between actors
as the most intense conflict of which they are capable: if the difference
in utilities were quite low, then one would expect little effort to
impose or resist. Because the actors have different utility functions,
it is entirely possible for the stakes to be very high for one actor
and very low for the other.

The proportional voting rule of equation \eqref{eq:Proportional_Voting_Between_Alternatives}
leads to the following effort by each actor.

\begin{equation}
\begin{array}{rclcc}
v_{i}(i:j) & = & c_{i}\left(U_{i}\left(\theta_{i}\right)-U_{i}\left(\theta_{j}\right)\right) & \geq & 0\\
v_{j}(i:j) & = & c_{j}\left(U_{j}\left(\theta_{i}\right)-U_{j}\left(\theta_{j}\right)\right) & \leq & 0
\end{array}
\end{equation}

Thus, the probability of i winning depend on both the relative capabilities
and relative stakes. We modify equation \eqref{eq:Basic-bilateral-victory-prob}
to use not the total capability available but only that amount justified
by the stakes, as follows:

\begin{equation}
\begin{array}{rcl}
p\left[i\succ j\right] & = & \frac{|v_{i}(i:j)|}{|v_{i}(i:j)|+|v_{j}(i:j)|}\\
 & = & \frac{v_{i}}{v_{i}+v_{j}}
\end{array}\label{eq:Scaled-bilateral-victory-prob}
\end{equation}

For this particular model, we will assume that the utility to an actor
i of a state where all other positions are $\theta_{j}$ is simply
the sum of the utilities of positions:%
\footnote{Other models might take the utility of a state to be the utility to
i of the CW of the state. While this is intuitively appealing, it
introduces much more complexity.%
}

\begin{equation}
\begin{array}{rcl}
S & = & \left\{ \theta_{j}\right\} \\
U_{i}(S) & = & \sum_{j}U_{i}\left(\theta_{j}\right)
\end{array}
\end{equation}

For brevity, we will omit terms which cancel. Suppose actor i considers
challenging actor j. Considered as a bilateral contest, only $\theta_{i}$
and $\theta_{j}$ could change, so all other terms will be omitted
from status quo utility:
\begin{equation}
U_{i}(S)=U_{i}\left(\theta_{i}\right)+U_{i}\left(\theta_{j}\right)
\end{equation}

The expected utility for party i in a contest against j rests on the
assumption that the loser must adopt the winner's position. If i defeats
j, then the new positions are $\theta'_{i}=\theta_{i}$ and $\theta'_{j}=\theta_{i}$;
the converse holds if j defeats i. Thus, the expected utility of the
outcome is as follows:

\begin{equation}
\begin{array}{rcl}
U_{i}(i:j) & = & \frac{v_{i}}{v_{i}+v_{j}}\left(U_{i}\left(\theta'_{i}\right)+U_{i}\left(\theta'_{j}\right)\right)+\frac{v_{j}}{v_{i}+v_{j}}\left(U_{i}\left(\theta''_{i}\right)+U_{i}\left(\theta''_{j}\right)\right)\\
 & = & \frac{v_{i}}{v_{i}+v_{j}}\left(U_{i}\left(\theta_{i}\right)+U_{i}\left(\theta_{i}\right)\right)+\frac{v_{j}}{v_{i}+v_{j}}\left(U_{i}\left(\theta_{j}\right)+U_{i}\left(\theta_{j}\right)\right)\\
 & = & 2\frac{v_{i}U_{i}\left(\theta_{i}\right)+v_{j}U_{i}\left(\theta_{j}\right)}{v_{i}+v_{j}}
\end{array}
\end{equation}

Comparing $U_{i}(i:j)$ to the status quo utility, we can see that
$i$'s incentive to initiate a one-on-one contest $j$ is the following:
\begin{equation}
U_{i}(i:j)-U_{i}(S)=\frac{v_{i}-v_{j}}{v_{i}+v_{j}}\left(U_{i}\left(\theta_{i}\right)-U_{i}\left(\theta_{j}\right)\right)\label{eq:Basic-Incentive-to-Challenge}
\end{equation}

Taken by itself, equation \eqref{eq:Basic-Incentive-to-Challenge}
implies that each actor has an incentive to initiate a contest which
depends on the interaction of capability and stakes. Notice that it
can be advantageous for a weak actor, $i$, to confront a strong actor,
$j$, if the stakes are small for the large actor yet high for the
weak actor: $c_{i}<c_{j}$ yet $v_{i}>v_{j}$ so $U_{i}(i:j)-U_{i}(S)>0$.
This dynamic is often cited as an explanation of how weak countries
can defeat strong countries, e.g. when the former fight without end
on their home territory and the latter easily tire of overseas expeditions.

However, all the third parties also have the opportunity to exert
influence, so contests cannot be expected to always remain bilateral
between just two actors.

\subsection{Third Party Support\label{sub:Third-Party-Support}}

As mentioned earlier, third parties are likely to gain or lose from
a contest and hence have an incentive to vote with $v_{k}(i:j)$.
Those supporting $i$ make up one coalition while those supporting
$j$ make another. In this way, we can view the multilateral contest
as a two-sided contest between two coalitions. The composition and
strength of the coalitions depend on the votes of third parties.

Third parties always have at least three options: support i, support
j, or abstain. Again, one would expect them to exert effort to support
the party whose victory is most favorable to them. We assume that
the third parties have limited rationality in that they look at the
first-order, local choices and consequences, without complex game-theoretic
consideration of others' choices (which might be based on similarly
complex consideration of others' choices, and so on.) 

We model third party support as uncommitted, semi-committed, or fully
committed.

\subsubsection{Support When Uncommitted }

In the uncommitted case, third parties are not committed to adopt
the position of whichever side they support, but the loser of the
bilateral contest is still forced to adopt the position of the victor.
The expected utility to k of supporting i would be simply the expected
value of the two outcomes. If i and k together defeat j, then i remains
at their initial position $\theta_{i}$, j must adopt position $\theta_{i}$,
and k can remain at position $\theta_{k}$. If i and k together lose
to j, then i must adopt $\theta_{j}$. When multiple actors form a
coalition, the probability of victory is determined by the strength
of each coalition, which is taken to be the sum of their individual
capabilities. Because positions for i, j, and k are involved, we carry
the terms of $U(S')$ and $U(S'')$ for all three positions:

\begin{equation}
\begin{array}{rcl}
U_{k}^{u}(i,k:j) & = & \frac{c_{i}+c_{k}}{c_{i}+c_{k}+c_{j}}\left(U_{k}\left(\theta'_{i}\right)+U_{k}\left(\theta'_{j}\right)+U_{k}\left(\theta'_{k}\right)\right)+\frac{c_{j}}{c_{i}+c_{k}+c_{j}}\left(U_{k}\left(\theta''_{i}\right)+U_{k}\left(\theta''_{j}\right)+U_{k}\left(\theta''_{k}\right)\right)\\
 & = & \frac{c_{i}+c_{k}}{c_{i}+c_{k}+c_{j}}\left(U_{k}\left(\theta_{i}\right)+U_{k}\left(\theta_{i}\right)+U_{k}\left(\theta_{k}\right)\right)+\frac{c_{j}}{c_{i}+c_{k}+c_{j}}\left(U_{k}\left(\theta_{j}\right)+U_{k}\left(\theta_{j}\right)+U_{k}\left(\theta_{k}\right)\right)
\end{array}
\end{equation}

Similarly for k supporting j:

\begin{equation}
U_{k}^{u}(j,k:i)=\frac{c_{i}}{c_{i}+c_{k}+c_{j}}\left(U_{k}\left(\theta_{i}\right)+U_{k}\left(\theta_{i}\right)+U_{k}\left(\theta_{k}\right)\right)+\frac{c_{j}+c_{k}}{c_{i}+c_{k}+c_{j}}\left(U_{k}\left(\theta_{j}\right)+U_{k}\left(\theta_{j}\right)+U_{k}\left(\theta_{k}\right)\right)
\end{equation}

As mentioned earlier, there are multiple voting rules discussed in
the literature, our prototypes use several, and expert judgment will
be required to determine which is most appropriate for any given case.
Because it is analytically clearer, we will present the results only
for proportional voting:

\begin{equation}
\begin{array}{rcl}
v_{k}^{u}(i:j) & = & w_{k}\left(U_{k}^{u}(i,k:j)-U_{k}^{u}(j,k:i)\right)\\
 & = & w_{k}\frac{2c_{k}}{c_{i}+c_{k}+c_{j}}\left(U_{k}\left(\theta_{i}\right)-U_{k}\left(\theta_{j}\right)\right)
\end{array}\label{eq:UC-third-party-votes}
\end{equation}

The particular case of equation \eqref{eq:UC-third-party-votes} (uncommitted,
proportional voting) is similar to the formula described in \cite{ForecastingPolEvents,PredPol},
albeit with different notation.

\subsubsection{Support When Semi-committed }

Of course, many small nations have learned that it matters greatly
whether their powerful allies win or lose, because they win or lose
with their patrons. In the semi-committed case, third party k can
keep $\theta_{k}$ if on the winning side but must adopt winner's
position otherwise:

\begin{equation}
U_{k}^{s}(i,k:j)=\frac{c_{i}+c_{k}}{c_{i}+c_{k}+c_{j}}\left(U_{k}\left(\theta_{i}\right)+U_{k}\left(\theta_{i}\right)+U_{k}\left(\theta_{k}\right)\right)+\frac{c_{j}}{c_{i}+c_{k}+c_{j}}\left(U_{k}\left(\theta_{j}\right)+U_{k}\left(\theta_{j}\right)+U_{k}\left(\theta_{j}\right)\right)
\end{equation}

Similarly for k semi-committed to j:

\begin{equation}
U_{k}^{s}(j,k:i)=\frac{c_{i}}{c_{i}+c_{k}+c_{j}}\left(U_{k}\left(\theta_{i}\right)+U_{k}\left(\theta_{i}\right)+U_{k}\left(\theta_{i}\right)\right)+\frac{c_{j}+c_{k}}{c_{i}+c_{k}+c_{j}}\left(U_{k}\left(\theta_{j}\right)+U_{k}\left(\theta_{j}\right)+U_{k}\left(\theta_{k}\right)\right)
\end{equation}

With proportional voting,

\begin{equation}
\begin{array}{rcl}
v_{k}^{s}(i:j) & = & w_{k}\left(U_{k}^{s}(i,k:j)-U_{k}^{s}(j,k:i)\right)\\
 & = & w_{k}\left[\frac{2c_{k}}{c_{i}+c_{k}+c_{j}}\left(U_{k}\left(\theta_{i}\right)-U_{k}\left(\theta_{j}\right)\right)+\frac{c_{i}\left(U_{k}\left(\theta_{k}\right)-U_{k}\left(\theta_{i}\right)\right)-c_{j}\left(U_{k}\left(\theta_{k}\right)-U_{k}\left(\theta_{j}\right)\right)}{c_{i}+c_{k}+c_{j}}\right]
\end{array}\label{eq:SC-third-party-votes}
\end{equation}

This is simply the uncommitted case, plus a term corresponding to
the danger of joining a losing coalition.

\subsubsection{Support When Fully Committed }

In the fully-committed case, third party k must always adopt the winner's
position:

\begin{equation}
U_{k}^{f}(i,k:j)=\frac{c_{i}+c_{k}}{c_{i}+c_{k}+c_{j}}\left(U_{k}\left(\theta_{i}\right)+U_{k}\left(\theta_{i}\right)+U_{k}\left(\theta_{i}\right)\right)+\frac{c_{j}}{c_{i}+c_{k}+c_{j}}\left(U_{k}\left(\theta_{j}\right)+U_{k}\left(\theta_{j}\right)+U_{k}\left(\theta_{j}\right)\right)
\end{equation}

Similarly for k fully committed to j:

\begin{equation}
U_{k}^{f}(j,k:i)=\frac{c_{i}}{c_{i}+c_{k}+c_{j}}\left(U_{k}\left(\theta_{i}\right)+U_{k}\left(\theta_{i}\right)+U_{k}\left(\theta_{i}\right)\right)+\frac{c_{j}+c_{k}}{c_{i}+c_{k}+c_{j}}\left(U_{k}\left(\theta_{j}\right)+U_{k}\left(\theta_{j}\right)+U_{k}\left(\theta_{j}\right)\right)
\end{equation}

With proportional voting,

\begin{equation}
\begin{array}{rcl}
v_{k}^{f}(i:j) & = & w_{k}\left(U_{k}^{f}(i,k:j)-U_{k}^{f}(j,k:i)\right)\\
 & = & w_{k}\frac{3c_{k}}{c_{i}+c_{k}+c_{j}}\left(U_{k}\left(\theta_{i}\right)-U_{k}\left(\theta_{j}\right)\right)
\end{array}\label{eq:FC-third-party-votes}
\end{equation}

Taken together, equations \eqref{eq:UC-third-party-votes} and \eqref{eq:FC-third-party-votes}
imply that modeling third parties as fully committed will increase
their effort by a factor of 3/2. As the behavior of the main parties
is the same in either case, the choice of voting rule can change the
result of equation \eqref{eq:Coalition-victory-prob} because the
purely bilateral contributions might dominate in the uncommitted case
while the larger third-party contributions might dominate in the fully
committed case.

Regardless of how committed the third parties might be if they support
one side or another, they always have the third option of abstaining
with $v_{k}(i:j)=0$. In that case, the utility to actor $k$ is just
their own expected utility of the bilateral contest:

\begin{equation}
U_{k}^{a}(i:j)=\frac{c_{i}}{c_{i}+c_{j}}\left(U_{k}\left(\theta_{i}\right)+U_{k}\left(\theta_{i}\right)+U_{k}\left(\theta_{k}\right)\right)+\frac{c_{j}}{c_{i}+c_{j}}\left(U_{k}\left(\theta_{j}\right)+U_{k}\left(\theta_{j}\right)+U_{k}\left(\theta_{k}\right)\right)\label{eq:Third-party-fully-committed-proportional}
\end{equation}

If $U_{k}^{a}(i:j)\geq U_{k}(j,k:i)$ and $U_{k}^{a}(i:j)\geq U_{k}(i,k:j)$
then the utility of abstaining is greater than the utility of supporting
either side, and that particular third party will abstain. This represents
the real-world situation of ``sitting on the fence'' until it becomes
clear that the dangers of full commitment or even semi-commitment
are worthwhile, i.e. until a ``strong horse'' emerges.

Again, equations like \eqref{eq:UC-third-party-votes}, \eqref{eq:SC-third-party-votes},
\eqref{eq:FC-third-party-votes}, \eqref{eq:Third-party-fully-committed-proportional}
do not actually appear in our prototype implementations. They implement
the general NPCE method, and use the abstract computation. However,
under each specific parameterization, the behavior of the abstract
method can be described by each specific equation, even though the
same abstract computation is used in each case. The situation is entirely
analogous to linear programming, where the general simplex algorithm
can be applied to data sets with many different structures. For some
structures (e.g. multi-commodity flow), there exists more specific
algebraic descriptions of how the general algorithm behaves with that
specific structure.

\subsection{Search for a Condorcet Winner\label{sub:Locating-a-CW}}

Much of the literature uses an abstract model inspired by elections:
the group simply selects one from a small set of predefined options.
The unidimensional issue set in the SPM is often characterized by
the integers from 0\% to 100\%, or even just the positions of a dozen
or so actors: it is easy to check each option for the CW. Even in
multidimensional SPM, the positions currently held by actors is a
small set that can be easily searched. With discrete, combinatorial
issue sets, one quickly encounters enormous sets of options: the ``combinatorial
explosion'' of options is unavoidable.

In infinite (or just astronomically large) issue sets, it is generally
impossible to analytically solve for the CW, so any effective procedure
- either for a modeling toolkit or a legislative process - must do
some sort of incremental exploration of the issue set, looking for
local improvements until no more can be found. Both hill-climbing
and genetic search are examples. Each alternates between two steps:
generating new local options, and selecting which to discard and which
to further improve. Because the complicated interactions of the generation
and search procedure are impossible to predict in detail (otherwise
analytic solution for the CW would be possible), the search for a
CW in a large combinatorial issue set is similar to the ``preference
construction'' process described in \cite{MABinChoiceStrat} and
discussed in \cite{CWMyth}:\emph{ }
\begin{quote}
\emph{Alternatives are processed in pairs, with the values of the
two alternatives compared on each attribute, and the alternative with
a majority of winning (better) attribute values is retained. The retained
option is then compared with the next alternative from the choice
set, and this process of pairwise comparison continues until all the
alternatives have been evaluated and one option remains.}
\end{quote}
This can be interpretted as a hill-climbing procedure to maximize
the $\omega$ function of equation \eqref{eq:CPT-proof}. The CPT
suggested that the observed convergence to definite legislative outcomes
can be explained by the existence of a CW, as well as by appeals to
institutional structures. Modified for a two-sided comparison process,
this could just as well describe the classic ``drift to war'' (see
section \eqref{sub:Normative-vs.-Descriptive}) where each side observes
the behavior of the other side and responds as seems best at the time,
the other side reacts similarly, and the process continues. Each comparison
of alternative responses can be seen as another one-on-one comparison
of options in an implicit Condorcet election, similar to the legislative
process of voting on successive proposed amendments.

In general, optimization algorithms can settle on different local
optima based on their starting point and search procedure; political
scientists often refer to the real-world manifestation of this phenomenon
as ``path dependence''. Therefore, researchers could pursue two
different paths: model the particular incremental, path-dependent
search process applicable to their particular problem, or try to design
a search algorithm which is likely to find the true global optimum
within reasonable time limits. However, the optimization issues involved
are not unique to the NPCE methodology, so they will not be discussed
here.

\section{Summary and Conclusions}

This paper has presented a mathematical framework that significantly
extends the very successful voting models inspired by the Median Voter
Theorem (MVT). Two broad classes of generalization are described.
The limitation of previous methods to representing outcomes via uni-dimensional
and continuous scales is generalized to allow discontinuous and multidimensional
spaces that can allow complex structured policies to be analyzed.
And by representing negotiation and bargaining as a non-deterministic
processes this approach will allow the exploration of possible outcomes
in situations that are sufficiently complex and uncertain that relying
on deterministic forecasts could be misleading.

Increased computational power means that simplifications previously
made for analytic tractability are no longer required. The combination
of the more general mathematical framework presented here and modern
computational resources now allow applications utilizing this framework
where earlier simpler formulations would not be suitable. Issues such
as the formulation of legislation, the formation of coalition governments,
the influence of stakeholder groups on direct parties to negotiations
and deliberations can now be analyzed using this framework. Further,
the more general approach can serve to avoid the hazard of excluding
important information from analyses in order to make them feasible.
Deterministic models have the potential to mislead by portraying best
estimate outcomes as certain, and the non-deterministic framework
described here can serve statistical properties of outcome distributions,
which can be important when there are a large number of complex policy
options.

\pagebreak{}

\end{document}